\documentstyle[12pt,epsf]{article} 

\def\abstract#1{\vskip 7mm 
        \begin{center}{\large Abstract}\par \smallskip 
                       \begin{minipage}[c]{12cm}
                        \small #1
                       \end{minipage}
        \end{center}
\smallskip 
\vspace{2mm}
}
\def\title#1{\begin{center}{\Large\bf #1}\end{center}}
\def\author#1{\vskip 5mm \begin{center}{#1}\end{center}}
\def\address#1{\begin{center}{\it #1}\end{center}}

\begin{document}

\def\odai#1{\vspace*{0.5cm} \fbox{ #1}\\}
\def\Mark#1{\vspace*{0.5cm} \fbox{{\bf #1}}\\}
\newtheorem{theorem}     {Theorem}
\newtheorem{lemma}       [theorem] {Lemma}
\newtheorem{proposition} [theorem] {Proposition}
\newtheorem{corollary}   [theorem] {Corollary}
\newtheorem{definition}  [theorem] {Definition}
\newtheorem{conjecture}  [theorem] {Conjecture}
\newtheorem{condition}   [theorem] {Condition}
\newtheorem{claim}       [theorem] {Claim}
\newenvironment{proof}{\noindent{\em Proof.}}%
{{\hspace*{1em}\hfill{$\Box$}}}
\newenvironment{proofof}[1]{\noindent{\em Proof of {#1}.}}%
{{\hspace*{1em} \hfill{$\Box$}}}
\newenvironment{remark}{\noindent{\bf Remark.}}{}


\begin{titlepage}
\hfill
\parbox{6cm}{{TIT/HEP-377/COSMO-85}} 

\hfill
\par
\vspace{5mm}
\title{\bf {Chronology Protection and Non-Naked Singularity} }
\vskip 7mm
\author{Kengo Maeda \footnote{e-mail:maeda@th.phys.titech.ac.jp; JSPS fellow} 
and Akihiro Ishibashi \footnote{e-mail:akihiro@th.phys.titech.ac.jp}}
\address{Department of Physics, Tokyo Institute of Technology, \\ 
Oh-Okayama Meguro-ku, Tokyo 152, Japan}  
\author{Makoto Narita \footnote{e-mail:narita@rikkyo.ac.jp}}
\address{Department of Physics, Rikkyo University, \\ 
  Nishi-Ikebukuro, Toshima, Tokyo 171, Japan}

\vskip 7mm 
\abstract{We test the chronology protection conjecture in classical general
relativity by investigating finitely vicious space-times. 
First we present singularity theorems in finitely vicious space-times 
by imposing some restrictions on the chronology violating sets. 
In the theorems we can refer to the location of an occurring singularity 
and do not assume any asymptotic conditions such as the existence of null 
infinities. 
Further introducing the concept of a {\it non-naked singularity}, 
we show that a restricted class of chronology violations cannot arise 
if all occurring singularities are the non-naked singularities.  
Our results suggest that the causal feature of the occurring 
singularities is the key to prevent the appearance of causality violation. 
}

\end{titlepage}

\section{Introduction} 
\label{I}  

Whether or not there are causality violating regions in the universe 
is an interesting problem in physics. 
Hawking has proposed the chronology protection conjecture 
which states that the laws of physics prevent closed timelike curves 
from appearing~\cite{H}. 
He investigated the instability of a Cauchy horizon caused 
by causality violation and 
showed that the renormalized stress-energy tensor 
always diverges at the Cauchy horizon by using quantum field theory. 
The instabilities of such Cauchy horizons have also been shown in 
this approach by several authors~\cite{TM}. 
On the other hand, Li et al.~\cite{Li} obtained the opposite result that 
the Cauchy horizon is stable against the vacuum fluctuations 
under some conditions. 
There also remains an unsettled problem that 
the back reaction of quantum effects on the geometry is difficult 
to take into account. 
At present, there seems no consensus on the quantum instability 
of the Cauchy horizon caused by causality violation. 
%

%
In the framework of classical general relativity, 
only a few attempts have so far been made at verifying 
the chronology protection conjecture. 
Hawking~\cite{H} considered a compactly generated Cauchy horizon 
that all the past-directed null geodesic generators of the Cauchy horizon 
are totally imprisoned in a compact set. He showed that such a Cauchy horizon 
cannot occur under the weak energy condition. 
Tipler provided a partial answer to the conjecture 
by showing that the creation of a closed timelike curve leads to 
singularity formation under the stronger energy condition than 
the usual energy conditions~\cite{T,T1}. 
Ori~\cite{Ori} discussed the relation 
between causality violation and the weak energy condition. 
In the context of testing the conjecture, 
one would like to examine the case that causality violation 
could arise in a finite region of space from a regular initial data 
by physical process such as gravitational collapse.  
In such a case space-time may be {\it finitely vicious}, 
whose notion was first introduced by Tipler~\cite{T1}. 
Space-times containing time machines are examples 
of finitely vicious space-times and 
many models of such space-times have been proposed. 
Morris et al.~\cite{KS} suggested a model of a time machine in a wormhole 
space-time by accelerating one mouth of the wormhole 
with respect to the other mouth. 
Deutsch and Politzer~\cite{DP} constructed a simple model of 
a time machine (the Deutsch-Politzer time machine) 
by identifications of spacelike line segments 
in two-dimensional Minkowski space-time. 
It is interesting that the null geodesic generators of the boundary of 
causality violating set in this model are not imprisoned in any compact sets, 
while those of a compactly generated Cauchy horizon are. 
%
The geometrical property of the Doutsch-Politzer time machine in 
asymptotically flat space-time is investigated by 
Chamblin et al.~\cite{CGS}.  

Most recently, Krasnikov~\cite{Kras} proposed a simple model of 
time machine by modifying the Deutsch-Politzer time machine. 
Surprisingly, there are no singularities associated with the chronology 
violation in this model even though the weak energy condition is satisfied. 
This model demonstrates that the weak energy condition does not 
necessarily prevent 
causality violations of the Deutsch-Politzer time machine type 
from occurring in classical general relativity; the chronology protection 
conjecture is violated even if this energy condition holds. 
%

In this paper, we test the chronology protection conjecture by 
investigating finitely vicious space-times under some reasonable conditions, 
especially, the strong energy condition. 
To begin with we define a class of finitely vicious space-times 
whose causal structures are similar to that of a space-time 
with the Deutsch-Politzer time machine. 
Next we show that there exists an inextendible and incomplete 
causal geodesic in the chronology violating region 
under physically reasonable conditions. 
To prove this, we do not assume the existence of any asymptotic regions. 
Introducing the notion of a {\it non-naked singularity}, 
we finally show that some classes of chronology violations 
which will be specified by a suitable boundary condition for the chronology 
violating set cannot arise if all occurring singularities 
are the non-naked singularities. 
In the next section, we consider a geometry of finitely vicious space-time.  
In Sec.~\ref{III}, we present our singularity theorems in finitely 
vicious space-times. 
On the bases of the results in Sec.~\ref{III}, we will present 
our main theorem in Sec.~\ref{IV}. 
Section~\ref{V} is devoted to summary and discussion.

\section{A geometry of chronology violation}
\label{II}

We shall consider space-times in which causality violating sets 
(chronology violating sets) are formed in gravitational collapse 
or ``manufactured'' as time machines by future technology. 
In such a space-time causality violation should begin in a finite region 
of space in the future of a partial Cauchy surface $S$. 
It is expected that associated with the causality violation 
a Cauchy horizon $H^{+}(S)$ with compact spatial section develops 
in the future of $S$. 
Such space-times are classified in a {\it finitely vicious} space-time 
by Tipler~\cite{T}, whose precise definition is given in the following. 
\\ 
{\it 
It is said that a space-time $(M,g)$ is finitely vicious 
if it has a hypersurface slicing $S(\tau)$ with the properties \\ 
(i) $S$ is one of the slices with $S(0)=S$; \\ 
(ii) there is a closed interval $[\tau_1,\tau_2]$ of the slice parameter
such that if $\tau\in [\tau_1,\tau_2]$, then 
$S(\tau)\cap{\tilde{D}}^{+}(S)$ is spacelike, and $S(\tau)\cap H^{+}(S)$  
is compact. Also, if $\tau_3,\,\tau_4$ are any two numbers in 
$[\tau_1,\tau_2]$ with $\tau_4\ge\tau_3$, then 
$S(\tau_4)\cap{\tilde{D}}^{+}(S)$ lies to the future of 
$S(\tau_3)\cap{\tilde{D}}^{+}(S)$; \\ 
(iii) Let $B$ be the region of space-time in ${\tilde{D}}^{+}(S)$ 
between $S(\tau_1)$ and $S(\tau_2)$ inclusive, and let $\gamma$ be any segment
of a generator of $H^{+}(S)$ with $\gamma\cap S(\tau_2)\neq\emptyset$ and 
$\gamma\subset B$. \\ 
Then $\gamma$ can be extended in $H^{+}(S)\cap B$ 
such that the extension intersects each $S(\tau)$ 
for $\tau\in [\tau_1,\tau_2]$ exactly once.  
} 

\begin{figure}[htbp]
 \centerline{\epsfxsize=10.0cm \epsfbox{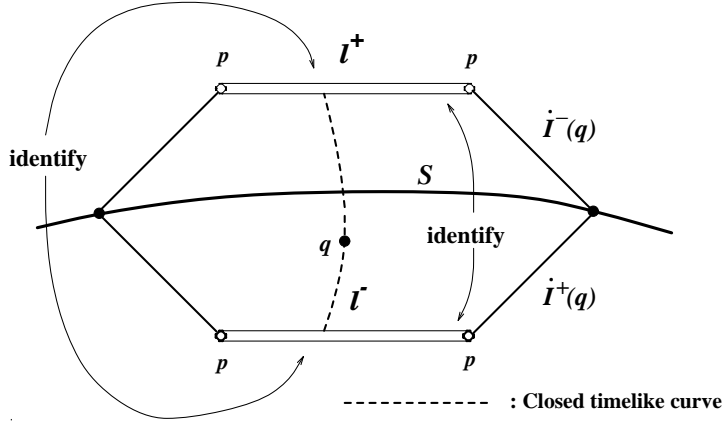}}
         \caption{The four-dimensional Deutsch-Politzer time machine $M$ is 
               shown. The hexagon is a chronology violating-region.  
               $S$ is a locally spacelike hypersurface. 
               Each point represents a closed spacelike two-surface $Q$. 
               The points $p$, which correspond to the points $(|t|=a,|x|=b)$ 
               in $I$, are removed. }
               \protect
\label{DP}
\end{figure}

There are some works on constructions of time machines~\cite{KS,DP,Kras} 
in finitely vicious space-times with non-trivial topology. 
The four-dimensional Deutsch-Politzer time machine~\cite{DP}, 
constructed below, is one of the simplest models of such time machines. 
Consider two-dimensional Minkowski space-time $I$ with metric    
\begin{eqnarray}
\label{eq-Min}
ds^2 = -dt^2+dx^2, 
\end{eqnarray}
and two horizontal line segments $l_{\pm} := \{ t,x | t = \pm a, |x|<b \} $ 
on $I$. Remove the points $(|t|=a,|x|=b)$ from $I$ and 
make cuts along $l_{\pm}$. Identifying the upper banks of $l_{+}$ and $l_{-}$ 
with the lower banks of $l_{-}$ and $l_{+}$, respectively, 
one can construct a two-dimensional space-time $I'$ containing 
the two-dimensional Deutsch-Politzer time machine. 
Then one can obtain the four-dimensional Deutsch-Politzer 
time machine as a product space-time $M = I'\times Q$, 
where $Q$ is a closed compact two-dimensional space, like $S^2$ or $T^2$. 
The causal structure of this space-time is schematically depicted 
in Fig.~\ref{DP}. 
 
The chronology violating-region, 
the interior of the hexagon in Fig.~\ref{DP}, 
is described as $I^{+}(q) \cap I^{-}(q)\, , q \in M$. 
The boundary of the chronology violating-region consists of portions of 
${\dot{I}}^{+}(q)$ and ${\dot{I}}^{-}(q)$. 
As easily seen in Fig.~\ref{DP}, the space-time is finitely vicious 
because there is a hypersurface $S$ whose intersection with 
${\dot{I}}^{+}(q)\cap{\dot{I}}^{-}(q)$ is compact. 

From the above observations, 
as a simple and reasonable class of finitely vicious space-times, 
we focus our attention on a finitely vicious space-time $(M,g)$ with 
a chronology violating set $V$ satisfying the following three conditions; \\ 
{\bf Conditions} \\ 
{\it 
1. there is a locally spacelike three-dimensional hypersurface $S$ 
(with no edge) such that 
$K := V\cap S$ has a compact closure $\overline{K}$ 
and the boundary $\dot{K}$ is equal to 
$ \dot{I}^{+}(q)\cap {\dot{I}}^{-}(q)$, \\ 
2. every closed causal curve in $\overline{V}$ is
a noncontractible curve which intersects $\overline{K}$, \\ 
3. $R_{ab}K^aK^b\ge 0$ for every non-spacelike vector $K^a$ 
and every causal geodesic with a tangent vector $K^a$ 
contains a point at which $K_{[a}R_{b]cd[e}{K_{f]}}K^cK^d \neq 0$. 
} 

Condition 1 is a boundary condition for the chronology violating set. 
Condition 2 is imposed for the simplicity of arguments since 
we consider closed causal curves associated with non-trivial
topology such as the topology in the Deutsch-Polizter time machines.  
Condition 2 ensures that the causality condition holds on the covering 
manifold $\hat{M}$ that will be considered in the following sections. 
Condition 3 means that the strong energy condition and 
the generic condition are satisfied. 

It should be remarked that in general 
the chronology violating set is described as the disjoint union 
of sets of the form $I^{+}(p) \cap I^{-}(p), \, p \in M$ 
(see Prop. 6.4.1 of Ref.~\cite{HE}). 
For simplicity, in the rest of this paper, we shall restrict our discussion 
to a single component of the chronology violating set 
without loss of generality and denote the component by $V$, 
unless there arises confusion.

\section{Singularity theorems}
\label{III}

In this section, we will investigate finitely vicious space-times 
satisfying conditions~$1,2,3$ and present theorems which state that 
there is a singularity in the causality violating-region.  
Before showing this, we shall explicitly define the statement 
$\lq\lq$ there is an incomplete causal geodesic curve (singularity) 
in a sub-region $N$ in a space-time" as follows.\\
\begin{definition}
\label{def-com}
Consider a submanifold $N$ of a space-time $(M,g)$.  
If there is an affinely parameterized causal geodesic 
$\gamma:[0,a) \rightarrow M$ which is inextendible in $M$ and 
incomplete at the value $a$ such that for each value $t\in [0,a)$ 
the point $\gamma(t)$ exists in $N$, 
then we can say that $N$ is causally geodesically incomplete.
\end{definition}

Now we shall show that $\overline{V}$ is causally geodesically 
incomplete without imposing any singularity conditions (Theorem~\ref{th}) 
to begin with. 
On the basis of Theorem~\ref{th}, we next present a theorem 
(Theorem~\ref{th1}) which states that $V$ is causally geodesically incomplete 
under the strong curvature singularity condition. 
To prove the theorems, for the space-time $(M,g)$ under consideration, 
we often use the covering manifold $\hat{M}$ defined as 
the set of all pairs $(p,[\lambda])$ where $[\lambda]$ is an 
equivalence class of curves from a locally spacelike hypersurface 
(with no edge) $S$ to $p \in M$ homotopic modulo $S$ and $p$ 
(see Ref.~\cite{HE}, p.~205). 
Let $\pi: \hat{M}\rightarrow M$ be a projection. Then, for any point $p$
of $M$, there exists an open neighborhood $U$ of $p$ such that $\pi^{-1}(U)$
is a collection of open sets of $\hat{M}$, each mapped diffeomorphically
onto $U$ by $\pi$. It is worth noting that 
each connected component of the image of $S$ is achronal and
homeomorphic to $S$ because of the construction of $\hat{M}$. 
For chronology violating set $V$, a set $\hat{V}$ 
is defined as the set of all points $\hat{q}\in\hat{M}$ 
such that $\pi(\hat{q})\in V$. 
Similarly, for a subset $N$ in $M$, $\hat{N}$ denotes hereafter 
the corresponding subsets $\pi^{-1}(N)$ in $\hat{M}$, 
unless otherwise explicitly stated. 

\subsection{Singularity Theorem I} 
\label{IIIA}

We present the following theorem. 

\begin{theorem}
\label{th}
If a finitely vicious space-time $(M,g)$ with a chronology 
violating set $V$ satisfies conditions~$1,2,3$, 
then $\overline{V}$ is causally geodesically incomplete.
\end{theorem}
To prove this theorem, we give the following 
proposition and lemmas under conditions~$1,2,3$, 
provided that $\overline{\hat{V}}$ is causally geodesically complete.    
\begin{proposition}
\label{pr-ret}
There is no causal curve which leaves $V$ and then returns to $V$.
\end{proposition}
\begin{proof}
Suppose that there was a causal curve $\gamma$ which left $V$ 
and reentered $V$, i.e., $\gamma$ had endpoints $p,r \in V$ 
and contained a point of $M-V$ between $p$ and $r$. 
Let us consider only the case that $\gamma$ was past-directed from $p$ 
to $r$ without loss of generality. 
Then there would be a point $s \in \gamma$ on $\dot{V}$ such that $p>s>r$. 
Since $V$ can be described as $I^{+}(p)\cap I^{-}(p)$, 
$s$ was in $\dot{I}^{+}(p)$ or $\dot{I}^{-}(p)$. 
Consider the case that $s\in{\dot{I}}^{-}(p)$. 
Because $p\in V$, there would be a closed timelike curve 
$\gamma_1$ through $p$. 
Since the causal curve $\gamma_1+\gamma$ is 
not a null geodesic, one could obtain a past-directed timelike curve 
$\gamma_2$ from $p$ to $s$ by deformation of $\gamma_1+\gamma$ 
following Prop.~4.5.10 of Ref.~\cite{HE}. 
This contradicts $s\in {\dot{I}}^{-}(p)$. 
Therefore $s\in {\dot{I}}^{+}(p)$. 
Because $r \in I^{+}(p)$ [recall that $r \in V = I^{+}(p) \cap I^{-}(p) $ ], 
there would be a past-directed causal curve $\gamma_3$ 
from $r$ to $p$. Then there was a past-directed causal curve $\gamma_4$ 
from $s$ to $p$ which consists of a segment of $\gamma$ from $s$ to $r$ 
and $\gamma_3$. 
Then, from Prop.~4.5.10 of Ref.~\cite{HE}, 
one could obtain a past-directed timelike curve which connects $s$ with $p$ 
by deformation of $\gamma_4 + \gamma_1$. 
This is impossible because $s\in{\dot{I}}^{+}(p)$. 
\end{proof}
\begin{lemma}
\label{pr-scc}
The strong causality condition holds in $\hat{V}$.
\end{lemma}
The proof is similar to that of Prop.~6.4.6 of Ref.~\cite{HE}. 

\begin{proof}
Suppose that the strong causality condition was violated 
at a point $\hat{p} \in \hat{V}$. 
Take a convex normal neighborhood $\hat{U}_p$ of $\hat{p} \in \hat{V}$ 
such that $\overline{\hat{U}_p} \subset \hat{V}$. 
Let $\hat{U}_n \subset \hat{U}_p$ 
be an infinite sequence of neighborhoods of $\hat{p}$ such that any 
neighborhood of $\hat{p}$ contains all the $\hat{U}_n$ for $n$ large enough. 
For each $\hat{U}_n$, there would be a future-directed causal curve 
$\hat{\lambda}_n$ in $\hat{V}$ which left $\hat{U}_n$ and then returned 
to $\hat{U}_n$. 
By Lemma 6.2.1 of Ref.~\cite{HE}, there would be an inextendible 
causal curve $\hat{\lambda}$ through $\hat{p}$ which was a limit curve 
of $\hat{\lambda}_n$. It follows from Prop.~\ref{pr-ret} that 
$\hat{\lambda} \subset \overline{\hat{V}}$ 
because, for each $n$, $\hat{\lambda}_n \subset \hat{V}$. 
Consider the case that $\hat{\lambda}$ was a null geodesic. 
Then $\hat{\lambda}$ could have a timelike separation, 
since $\overline{\hat{V}}$ was null geodesically complete 
and the generic condition holds by condition~3. 
In the case that $\hat{\lambda}$ was not a null geodesic, 
$\hat{\lambda}$ could also have a timelike separation. 
Thus for any case one could join up some $\hat{\lambda}_n$ to give 
a closed causal curve $\hat{\lambda'}_n$.  $\hat{\lambda'}_n$ was 
in $\hat{V}$ by Prop.~\ref{pr-ret}(recall that each 
$\hat{\lambda}_n$ is in $\hat{V}$ and therefore the closed causal 
curve $\hat{\lambda'}_n$ contains a point in $\hat{V}$).  
This is impossible because the covering manifold $\hat{M}$ contains 
no closed causal curves in $\overline{\hat{V}}$ by the condition~2.
\end{proof}
\begin{lemma}
\label{lem-ch}
Let $\hat{K}$ be one of the connected components of $\pi^{-1}(\overline K)$ 
and consider a future (past) Cauchy horizon $H^{+}(\hat{K})$ 
[resp. $H^{-}(\hat{K})$]. 
If $H^{+}(\hat{K})\cap\hat{V}$ 
[resp. $ H^{-}(\hat{K})\cap\hat{V}$] 
is non-empty, 
then each past-(future-)directed null geodesic generator through a point of 
$H^{+}(\hat{K})\cap\hat{V}$ 
[resp. $H^{-}(\hat{K})\cap\hat{V}$]
has no past (future) endpoint and remains in $\hat{V}$. 
\end{lemma}
%
%
\begin{proof}
Let us consider only $H^{+}(\hat{K})$ without loss of generality. 
Let $\hat{\xi}$ be a past-directed null geodesic generator 
through a point $\hat{r}$ of $H^{+}(\hat{K})\cap\hat{V}$. 
Suppose that $\hat{\xi}$ had a past endpoint $\hat{s}$, which was in   
${\rm edge}(\hat{K})$. 
By condition~1, $s = \pi(\hat{s})$ was 
in $\dot{I}^{+}(r,M)\cap\dot{I}^{-}(r,M)$, $r = \pi (\hat{r}) \in V$. 
Then there would be a past-directed causal curve 
$\xi = \pi(\hat{\xi}) \subset M$ from $r$ to $s$. 
This is impossible by the same argument in the proof of Prop.~\ref{pr-ret} 
since $s$ was in ${\dot{I}}^{-}(r,M)$. 
Therefore each null geodesic generator through a point of 
$H^{+}(\hat{K})\cap\hat{V}$ is past-inextendible. 

Suppose that $\hat{\xi}$ left $\hat{V}$. Let $\hat{t}$ be a point of 
$(\hat{M}-\hat{V}) \cap\hat{\xi}$ and $\hat{U}_t$ an arbitrary 
small neighborhood of $\hat{t}$. 
Since $\hat{t}$ would be in $H^{+}(\hat{K})$, 
$I^{-}(\hat{t})\cap\hat{U}_t\cap {\rm int}(D^{+}(\hat{K}))$ 
is non-empty. Take a point ${\hat{t}}'$ in 
$I^{-}(\hat{t})\cap\hat{U}_t\cap {\rm int}(D^{+}(\hat{K}))$. 
Then one could get a past-directed causal curve $\hat{\mu}$ from
$\hat{r}\in\hat{V}$ to $(\hat{K}-{\rm edge}(\hat{K})) \subset \hat{V}$ 
through $\hat{t}$ and $\hat{t}'$ because 
$\hat{t}'\in {\rm int}(D^{+}(\hat{K}))$. 
The existence of $\hat{\mu}$ would imply that 
a causal curve $\mu = \pi(\hat{\mu})$ left $V$ and 
then returned to $V$ in $M$. 
This is impossible by Prop.~\ref{pr-ret}. 
Therefore each past-directed null geodesic generator through a point of 
$H^{+}(\hat{K})\cap\hat{V}$ remains in $\hat{V}$.
\end{proof}

\begin{lemma}
\label{lem-gamma1} 
There exists a future-inextendible timelike curve $\hat{\gamma}$ 
in ${\rm int}(D^{+}(\hat{K}))$ and 
$J^{-}(\hat{\gamma})\cap J^{+}(\hat{K})$ is in 
$\hat{V}$.
\end{lemma} 
\begin{proof}
There exists a closed timelike curve $\gamma'$ through a point of $K$ 
because $K \subset V$. Then, in the covering manifold $\hat{M}$, 
$\hat{\gamma'} = \pi^{-1}(\gamma')$ is a future-inextendible 
timelike curve from a point of ${\rm int} (\hat{K})$. 

If $\hat{\gamma'}$ does not intersect $H^{+}(\hat{K})$, 
$\hat{\gamma'}$ is the desired future-inextendible timelike curve 
contained in ${\rm int}(D^{+}(\hat{K})) \cap \hat{V}$. 

Consider the case that $\hat{\gamma'}$ intersects $H^{+}(\hat{K})$. 
Denote the intersection point by $\hat{p}(\in\hat{V})$ 
and $I^{-}(\hat{p})\cap\hat{K}$ with a compact closure  
by $\hat{\Sigma}$, respectively.  
$[I^{-}(\hat{p})\cap J^{+}(\hat{K})] 
\subset D^{+}(\hat{K})$ 
because $\hat{p}\in (H^{+}(\hat{K})-\hat{K})$. This means that   
$[I^{-}(\hat{p})\cap J^{+}(\hat{K})]\subset
D^{+}(\overline{\hat{\Sigma}})$ by definition of $\overline{\hat{\Sigma}}$.   
The past-directed null geodesic generator $\hat{\xi}_p$ 
of $H^{+}(\hat{K})$ through $\hat{p}$ is a past-inextendible 
curve in $\hat{V}$ which does not intersect 
$\overline{\hat{\Sigma}}$ by Lemma~\ref{lem-ch}. 
Then $\hat{\xi}_p$ is also the null geodesic generator 
of $H^{+}(\overline{\hat{\Sigma}})\cap\hat{V}$ because 
$\hat{\xi}_p\in
[\overline{I^{-}(\hat{p})}\cap J^{+}(\hat{K})]
\subset\overline{D^{+}(\overline{\hat{\Sigma}})}$.   

Suppose that $H^{+}(\overline{\hat{\Sigma}})$  
was compact and hence it could be covered by a finite number 
of local causality neighborhoods $\hat{U}_i$. 
Since $\hat{\xi}_p 
\subset H^{+}(\overline{\hat{\Sigma}})\cap\hat{V}$ 
one could take a subcollection $\{\hat{U}_m\}$ 
of the neighborhoods $\hat{U}_i$ 
such that each $\overline{\hat{U}_m}$ was contained in $\hat{V}$ 
and $\hat{\xi}_p$ remained in the compact subset 
$\cup_{m} \overline{\hat{U}_m}$ of $\cup_{i}\overline{\hat{U}_i}$. 
This is impossible by Prop. 6.4.7 of Ref.~\cite{HE} 
because the strong causality condition holds on each 
$\overline{\hat{U}_m}(\subset\hat{V})$ by Lemma~\ref{pr-scc}.   
Therefore $H^{+}(\overline{\hat{\Sigma}})$ 
is non-compact. Following the proof of Corollary on page 268 
of Ref.~\cite{HE}, put a timelike vector field on $\hat{M}$. 
Because $\overline{\hat{K}}(\subset\hat{S})$ is achronal, 
$\overline{\hat{\Sigma}}$ is also achronal. 
Then, if every integral curve of the vector field which intersected 
$\overline{\hat{\Sigma}}$ also intersected 
$H^{+}(\overline{\hat{\Sigma}})$, 
they would define a continuous one-one mapping of $\overline{\hat{\Sigma}}$ 
onto
$H^{+}(\overline{\hat{\Sigma}})$ and hence 
$H^{+}(\overline{\hat{\Sigma}})$ would be compact. 
This is a contradiction.
Therefore there is a future-directed inextendible timelike curve
$\hat{\gamma}$ in ${\rm int}(D^{+}(\overline{\hat{\Sigma}}))$. 
$\hat{\gamma}$ does not intersect 
${\dot{I}}^{-}(\hat{p})\cap J^{+}(\hat{K})$, otherwise there would be
a point $\hat{s}\in\hat{\gamma}$ such that 
$(J^{-}(\hat{s})\cap\hat{K})\not\subset\overline{\hat{\Sigma}}$.    
Thus $\hat{\gamma}$ is contained in 
${\rm int}(D^{+}(\overline{\hat{\Sigma}}))\cap J^{-}(\hat{p})$ 
and also in ${\rm int}(D^{+}(\hat{K})) \cap J^{-}(\hat{p})$.

To prove that $J^{-}(\hat{\gamma}) \cap J^{+}(\hat{K})$ 
is also in $\hat{V}$, it is enough to show that 
$J^{-}(\hat{p})\cap J^{+}(\hat{K})$ 
does not contain any point of $\hat{M}-\hat{V}$ because 
$J^{-}(\hat{\gamma})\cap J^{+}(\hat{K})\subset 
J^{-}(\hat{p})\cap J^{+}(\hat{K})$. 
If there was a point of $\hat{M}-\hat{V}$ 
in $J^{-}(\hat{p})\cap J^{+}(\hat{K})$, 
there would be a past-directed causal curve from $\hat{p}$ 
to a point $\hat{r} \in\overline{\hat{\Sigma}}\cap\hat{V}$ 
through a point of $\hat{M}-\hat{V}$ 
(there is no past-directed causal curve from $\hat{p}$ to a point of 
{\rm edge}$(\hat{K})$ as $\hat{\xi}$ in the proof of Lemma~\ref{lem-ch}). 
This is impossible by Prop.~\ref{pr-ret} because $\hat{p}$ and $\hat{r}$ 
was in $\hat{V}$. 

In the case that $\hat{\gamma}'$ does not intersect 
$H^+(\hat{K})$, it also can be shown that 
$J^-(\hat{\gamma}') \cap J^+(\hat{K})$ is in $\hat{V}$. 
\end{proof}
\begin{lemma}
\label{lem-E} 
Let us suppose that there exists a past-inextendible timelike curve 
$\hat{\lambda}\in\hat{V}$ through a point of ${\rm int}(\hat{T})$, 
where $\hat{T} := \overline{J^{-}(\hat{\gamma})}\cap\hat{K}$
and $\hat{\gamma}$ is the future-inextendible timelike 
curve obtained in Lemma~\ref{lem-gamma1}.  
If $\hat{\lambda}$ intersects 
$H^{-}(\dot{J}^{-}(\hat{T}))$ 
at a point $\hat{p}(\in\hat{V})$,
each null geodesic generator of 
$[I^{+}(\hat{p})\cap\dot{J}^{-}(\hat{T})]$
is contained in $E^{-}(\hat{T})\cap\hat{V}$.    
\end{lemma}
\begin{proof}
Consider an infinitesimally small 
neighborhood $\hat{U}$ of $\hat{p}$ and denote a set 
$[\hat{U}\cap I^{+}(\hat{p})]$ by $\hat{C}$. 
Because $\hat{p}\in 
[H^{-}(\dot{J}^{-}(\hat{T}))-\dot{J}^{-}(\hat{T})]$,
$\hat{C}\subset{\rm int}[D^{-}(\dot{J}^{-}(\hat{T}))]$. 
Thus, every future-inextendible causal curve from any point
of $\hat{C}$ intersects ${\dot{J}}^{-}(\hat{T})$ and also 
$I^{+}({\dot{J}}^{-}(\hat{T}))$ by Lemma 6.6.4 of Ref.~\cite{HE}. 
Suppose that there was a future-inextendible null geodesic   
generator of $J^{+}(\hat{C})\cap
\dot{J}^{-}(\hat{T})$. This means that 
there was a future-inextendible causal curve from a point 
of $\hat{C}$ and remained in $\dot{J}^{-}(\hat{T})$.  
This is a contradiction.
Therefore 
every null geodesic generator of $J^{+}(\hat{C})\cap
\dot{J}^{-}(\hat{T})$ has a future endpoint in $\hat{T}$,      
which means that $[J^{+}(\hat{C})\cap  
\dot{J}^{-}(\hat{T})]\subset J^{-}(\hat{T})$
but $\not\subset I^{-}(\hat{T})$.
By definition of $E^{-}(\hat{T}):=J^{-}(\hat{T})-I^{-}(\hat{T})$,  
each null geodesic generator of
$[J^{+}(\hat{C})\cap  
\dot{J}^{-}(\hat{T})]\subset E^{-}(\hat{T})$ 
and hence $[I^{+}(\hat{p})\cap  
\dot{J}^{-}(\hat{T})]\subset E^{-}(\hat{T})$. 

Denote $[I^{+}(\hat{p})\cap  
\dot{J}^{-}(\hat{T})]$ by $\hat{\Sigma}$ and the null
geodesic generator of $\hat{\Sigma}$ by $\hat{\eta}$. 
The future endpoint of $\hat{\eta}$, $\hat{T}\cap\hat{\eta}$, is in 
${\rm int}(\hat{K})$ (It is impossible that the future endpoint of 
$\hat{\eta}$ is in $\dot{\hat{K}}$ by the facts that  
$\pi({\hat{p}})$ is in $V$ and 
$\dot{K}=\dot{I}^{+}(\pi{(\hat{p})}) \cap \dot{I}^{-}(\pi{(\hat{p}}))$ 
as discussed in Lemma~\ref{lem-ch}). 
Then it follows from Prop.~\ref{pr-ret} that 
every null geodesic $\hat{\eta}$ is contained in $\hat{V}$ 
because $\hat{p}$ and ${\rm int}(\hat{K})$ are contained in $\hat{V}$. 

\end{proof}

\begin{lemma}
\label{lem-gamma2}
There is a future and past-inextendible timelike curve in 
${\rm int}[D(E^-(\hat{T}))] \cap \hat{V}$. 
%
%
\end{lemma}
\begin{proof}
We have only to show that there is a past-inextendible timelike 
curve $\hat{\lambda}$ in ${\rm int}[D^{-}(E^-(\hat{T}))]$
because of Lemma~\ref{lem-gamma1}. 

There is a closed timelike curve $\lambda$ through a point of 
${\rm int}(\pi(\hat{T}))$ because ${\rm int}(\pi(\hat{T})) \subset V$ . 
This means that there exists a past-inextendible timelike curve
$\hat{\lambda} = \pi^{-1}(\lambda)$ in $\hat{V}$ from 
a point of ${\rm int}(\hat{T})$ because $\hat{\lambda}$ is inextendible.  
Let us consider a Cauchy development 
$D^{-}({\dot{J}}^{-}(\hat{T}))$. \\
\\
(a) Assume that $\hat{\lambda}$ does not intersect
$H^{-}({\dot{J}}^{-}(\hat{T}))$.  
Each point of $\hat{\lambda}$ is in ${\rm int}
[D^{-}({\dot{J}}^{-}(\hat{T}))]$ and hence 
every future-inextendible causal curve from $\hat{\lambda}$ 
intersects ${\dot{J}}^{-}(\hat{T})$ and also 
$I^{+}({\dot{J}}^{-}(\hat{T}))$. 
Then $[J^{+}(\hat{\lambda})\cap 
{\dot{J}}^{-}(\hat{T})]\subset E^{-}(\hat{T})$    
as shown in the proof of the previous lemma and hence 
$\hat{\lambda}$ is a desired curve.\\  
\\
(b) Assume that $\hat{\lambda}$ intersects 
$H^{-}({\dot{J}}^{-}(\hat{T}))$. 
Denote the intersection point and the future-directed 
null geodesic generator of $H^{-}({\dot{J}}^{-}(\hat{T}))$ 
through the point by $\hat{p}$ and $\hat{\xi}_p$, respectively. 
$\hat{\xi}_p $ is future-inextendible by Prop.6.5.2 
and Prop.6.5.3 of Ref.~\cite{HE}. 
Let us consider a set   
$\hat\Sigma: =[I^{+}(\hat{p})\cap\dot{J}^{-}(\hat{T})]$.
By Lemma~\ref{lem-E}, each null geodesic 
generator of $\hat\Sigma$ is contained in 
$E^{-}(\hat{T})\cap\hat{V}$.    
If $\overline{\hat{\Sigma}}$ was non-compact, 
there would be a past-inextendible null geodesic generator 
$\hat{\eta}$ of $E^{-}(\hat{T})\cap\hat{\Sigma}$. 
This is impossible because the null geodesic generator $\hat{\eta}$ 
has a past endpoint by condition~3, 
otherwise one could get a past incomplete null geodesic $\hat{\eta}$ 
in $\hat{V}$ or a future incomplete null geodesic generator of 
${\dot{J}}^{-}(\hat{\gamma})$ in $\overline{\hat{V}}$ since 
$J^-(\hat{\gamma})\cap J^+(\hat{K})\subset\hat{V}$
by Lemma~\ref{lem-gamma1}. 
Therefore $\overline{\hat{\Sigma}}$ is compact.\\ 
(b-I) Assume that $\hat{\xi}_p$ does not intersect 
$E^{-}(\hat{T})\cap\overline{\hat{\Sigma}}$. It can be shown 
that $\hat{\xi}_p$ remains in $\hat{V}$ 
by the same argument in Lemma~\ref{lem-ch}. 
Then $H^{-}(\overline{\hat{\Sigma}})$ 
is non-compact by the facts that 
$\hat{\xi}_p \subset (\hat{V}\cap H^{-}(\overline{\hat{\Sigma}}))$ 
and the strong causality condition holds in $\hat{V}$ 
as shown in the proof of Lemma~\ref{lem-gamma1}. 
Then one can get a past-inextendible timelike curve 
$\hat{\lambda}$ in 
{\rm int}$[D^{-}(E^{-}(\hat{T}))]\cap\hat{V}$ 
because of the compactness of $\overline{\hat{\Sigma}}$ 
as in Lemma~\ref{lem-gamma1}.\\ 
(b-II) Assume that $\hat{\xi}_p$ intersects 
$E^{-}(\hat{T})\cap\overline{\hat{\Sigma}}$. $\hat{\xi}_p$ is a null 
geodesic generator of ${\dot{J}}^{-}(\hat{T})-E^{-}(\hat{T})$
because $\hat{\xi}_p$ is future-inextendible 
(remind that the future-directed null geodesic generators of  
${\dot{J}}^{-}(\hat{T})-E^{-}(\hat{T})$ are inextendible
and also $\hat{\xi}_p$ and ${\dot{J}}^{-}(\hat{T})$ are achronal). 
Denote an intersection point of 
$\hat{\xi}_p\cap [E^{-}(\hat{T})\cap\overline{\hat{\Sigma}}]$ 
by $\hat{q} \, (\in\hat{V})$ and take an arbitrary small neighborhood 
$\hat{U}_q$ of $\hat{q}$ such that $\overline{\hat{U}_q} \subset \hat{V}$. 
Since $\hat{q}\in [{\dot{J}}^{-}(\hat{T})\cap 
H^{-}({\dot{J}}^{-}(\hat{T}))]$, 
there is a future-directed causal curve $\hat{\mu}_0$ from a point 
$\hat{q}_0 \, (\in\hat{V})$ of $J^{-}(\hat{T})
\cap[\hat{U}_q-D^{-}({\dot{J}}^{-}(\hat{T}))]$ 
to a point of $\hat{T} \cap \hat{V}$ 
($\hat{\mu}_0$ cannot have a future endpoint in $\dot{\hat{K}}$ 
as previously discussed $\hat{\eta}$ in Lemma~\ref{lem-E}).  
Then from Prop.~\ref{pr-ret}, it follows that 
$\hat{\mu}_0$ is contained in $\hat{V}$. 
Consider a collection of local causality neighborhoods $\hat{U}_\alpha$ 
covering $\hat{M}$ as a locally finite atlas and its subcollection 
$\{ \hat{U}_i \}$ which covers $H^{-}(E^{-}(\hat{T}))$. 
As discussed in Lemma~8.2.1 of Ref.~\cite{HE}, 
$\hat{\mu}_0$ intersects $H^{-}(E^{-}(\hat{T}))$ at a 
point $\hat{r}_1\in\hat{V}$.
Then one can take a neighborhood $\hat{U}_1$ of $\hat{r}_1$ such that
$\overline{\hat{U}_1}$ also is in $\hat{V}$. 
Let $\hat{p}_1$ be a point of $J^{-}(\hat{T}) \cap
[\hat{U}_1-D^{-}({\dot{J}}^{-}(\hat{T}))]$. 
There is a future-inextendible causal curve $\hat{\lambda}_1$ from $\hat{p}_1$ 
which does not intersect either 
${\dot{J}}^{-}(\hat{T})$ or $D^{-}(E^{-}(\hat{T}))$. 
Let $\hat{q}_1$ be a point on $\hat{\lambda}_1$ not in $\overline{\hat{U}_1}$. 
Since $\hat{q}_1\in J^{-}(\hat{T})$ and $\hat{p}_1\in\hat{V}$, 
there is a future-directed causal curve $\hat{\mu}_1$ in $\hat{V}$ 
which connects $\hat{q}_1$ with a point of $\hat{T}\cap\hat{V}$. 
By Lemma~\ref{pr-scc} this curve intersects 
$H^{-}(E^{-}(\hat{T}))$ at a point $\hat{r}_2 \in \hat{V}$ 
not in $\overline{\hat{U}_1}$. 
One can take a neighborhood $\hat{U}_2$ of $\hat{r}_2$ such that 
$\overline{\hat{U}_2}$ also is in $\hat{V}$. 
Thus one obtain an infinite sequence of points $\hat{r}_n \in \hat{V}$ 
such that any subsequence of $\{\hat{r}_n\}$ does not converge to any point 
in $\hat{M}$ by the construction of $\{\hat{r}_n\}$. 

Let us consider a future-directed null geodesic generator 
$\hat{\alpha}_n\in H^{-}(E^{-}(\hat{T}))$ 
from $\hat{r}_n$ $(\hat{\alpha}_m\neq\hat{\alpha}_{m'}$ if $m \neq m')$. 
There are now two different situations to be 
taken into account;\\ 
(b-II-i) Every $\hat{\alpha}_m$ does not have a future 
endpoint $\hat{s}_m$. 
There is a future-inextendible null geodesic generator 
$\hat{\alpha}_{m_o}$ of $H^{-}(E^{-}(\hat{T})) \cap \hat{V}$. 
Then one can obtain a past-inextendible timelike curve $\hat{\lambda}$ 
in $\hat{V}\cap {\rm int}[D^{-}(E^{-}(\hat{T}))]$ 
by the same way that we obtained it in the case (b-I).\\ 
(b-II-ii) Every $\hat{\alpha}_m$ has a future endpoint $\hat{s}_m$.   
One can get an infinite sequence of points 
$\hat{s}_m \in [{\rm edge}(E^{-}(\hat{T}))\cap\hat{V}]$ 
because $\hat{\alpha}_m \subset \hat{V}$ as $\hat{\mu}_n \subset \hat{V}$. 
Let $\hat{\beta}_m$ be a null geodesic generator of 
$E^{-}(\hat{T}) \cap \hat{V}$ whose past endpoint is $\hat{s}_m$. 
Take a sequence of future endpoints $\hat{t}_m \, (\in (\hat{T}\cap\hat{V}))$ 
of $\hat{\beta}_m$. 
From the compactness of $\hat{T}$ one can take a subsequence 
$\{ \hat{t}_l\}$ of $\{ \hat{t}_m\}$ which converges to a point $\hat{t}$ 
of $ \hat{T}\cap\overline{\hat{V}}$. 
By Lemma.~6.2.1 of Ref.~\cite{HE} there exists a limit curve $\hat{\beta}$ 
through a point $\hat{t}$ which is also a null geodesic generator 
of $E^{-}(\hat{T})$.    
If $\{\hat{s}_l\}$ did not converge, 
$\hat{\beta}$ would be past incomplete or there was a future incomplete 
null geodesic generator of $\dot{J}^{-}(\hat{\gamma})$ in 
$\overline{\hat{V}}$. 
Thus, $\hat{\beta}$ has a point $\hat{s}$ 
in ${\rm edge}(E^{-}(\hat{T}))$ which is a limit point 
of $\{\hat{s}_l \}$.
Because each null geodesic generator $\hat{\beta}_l$ is contained 
in $\hat{V}$, the segment of the limit curve $\hat{\beta}$ 
from $\hat{s}$ to $\hat{t}$ is contained in $\overline{\hat{V}}$. 
Consider an infinite sequence of null geodesics 
$\hat{\alpha}'_l $ such that each $\hat{\alpha}'_l ( \supset \hat{\alpha}_l)$ 
is past-inextendible in $\hat{M}$ and passes through two points $\hat{s}_l$ 
and $\hat{r}_l$. 
By Lemma 6.2.1 of Ref.~\cite{HE} through $\hat{s}$ 
there exists a limit null geodesic curve $\hat{\alpha}'$ 
which is past-inextendible. 
Since each $\hat{\alpha}_l$ is a null geodesic generator
of $H^{-}(E^{-}(\hat{T}))$, a portion of $\hat{\alpha}'$ 
is a null geodesic generator $\hat{\alpha}$ 
of $H^{-}(E^{-}(\hat{T}))$ from $\hat{s}$. 
Suppose that $\hat{\alpha}$ had a past endpoint 
in $H^{-}(E^{-}(\hat{T}))$.
Then one could take a point $\hat{a}$ on 
$\hat{\alpha'}\cap[\hat{M}-H^{-}(E^{-}(\hat{T}))]$ 
and the neighborhood $\hat{U}_a$ of $\hat{a}$ 
such that $\overline{\hat{U}_a} \subset 
[\hat{M}-H^{-}(E^{-}(\hat{T}))]$. 
Since $\hat{a}$ was a limit point of $\{\hat{\alpha}'_l\}$, 
one could take a point $\hat{a}_l (< \hat{r}_l)$ 
in $\hat{\alpha}'_l\cap\hat{U}_a$ for $l$ large enough. 
Then a subsequence of $\{\hat{r}_l\}$ would also converge to a point 
$\hat{b} \, (\hat{s}>\hat{b}>\hat{a})$ of $\hat{\alpha}$. 
This contradicts the fact that any subsequence of $\{\hat{r}_l\}$ 
has no limit point as shown above. 
Therefore $\hat{\alpha}(=\hat{\alpha}')$ is past-inextendible in 
$H^{-}(E^{-}(\hat{T}))$. 
Because $\hat{\alpha}_l \subset \hat{V}$, 
$\hat{\alpha}$ is contained in $\overline{\hat{V}}$. 
Obviously there is a past-inextendible timelike curve 
$\hat{\lambda}$ from ${\rm int} (\hat{T})$ in 
$I^{+}(\hat{\alpha})\cap 
{\rm int}[D^{-}(E^{-}(\hat{T}))]$. 
Then it can be also shown that $\hat{\lambda} \subset \hat{V}$ 
from Prop.~\ref{pr-ret} as in the case of $\hat{\mu}_0 \subset \hat{V}$. 
Thus, the proof of the case (b-II) is complete.
%

\end{proof}
\\ 

\begin{proofof}{theorem~\ref{th}} 
By Lemma~\ref{lem-gamma2}, there is a future and past-inextendible timelike
curve $\hat{\rho}$ in 
${\rm int}[D(E^{-}(\hat{T}))] \cap \hat{V}$. 
Then, following the proof of the theorem of Hawking and Penrose 
(see Ref.~\cite{HE}, p.~269), applied to $\hat{\rho}$, 
one can complete the proof of Theorem~\ref{th}. 

Take infinite sequences of points $\hat{a}_n$, $\hat{b}_n$ 
on $\hat{\rho}$ such that; \\  
(I) $\hat{a}_{n+1} \in I^{-}(\hat{a}_n)$, 
$(\hat{b}_{n+1}\in I^{+}(\hat{b}_n))$,  \\
(II) no compact segment of $\hat{\rho}$ contains more than a finite number
of the $\hat{a}_n$ $(\hat{b}_n)$, \\
(III) $\hat{b}_1\in I^{+}(\hat{a}_1)$. \\ 
Then there would be a timelike geodesic $\hat{\mu}_n$ of maximum length 
between $\hat{a}_n$ and $\hat{b}_n$ since $\hat{\rho}$ was contained 
in the globally hyperbolic set 
${\rm int}[D(E^{-}(\hat{T}))] \cap \hat{V}$. 
Because $\hat{T}$ is compact, there would be an inextendible
causal geodesic $\hat{\mu}$ 
in $J^{-}(\hat{\rho})\cap J^{+}(\hat{\rho})(\subset\hat{V})$. 
If $\hat{\mu}$ was future and past complete in $\hat{V}$,  
there would be conjugate points $\hat{x}$ and $\hat{y}$ on $\hat{\mu}$ 
with $\hat{y}\in I^{+}(\hat{x})$. This is a contradiction.
\end{proofof}
%
\\

We also have the following corollary.

\begin{corollary}
If the causality violating-region $\overline{V}$ which satisfies the 
conditions 1 and 2 is causally geodesically complete, the strong energy 
condition or the generic condition is violated. 
\hfill$\Box$
\end{corollary}

It is remarkable that, to prove Theorem~\ref{th}, any asymptotic conditions 
such as the existence of null infinities are not imposed.  

\subsection{Singularity Theorem II} 
\label{IIIB}

We have seen that $\overline{V}$ is causally geodesically incomplete; 
singularities are formed in $\overline{V}$. 
If a space-time singularity is caused by a physically realistic
process such as gravitational collapse, the curvatures become 
unboundedly large near the singularity and consequently 
all objects falling into it are crushed to be zero volume. 
This is conjectured by Tipler et al.~\cite{TCE} 
and independently by Kr\'olak~\cite{K'}. 
They introduced the notion of a 
{\it strong curvature singularity}~\cite{T2,K}. 
Let us consider an incomplete timelike (null) geodesic 
terminating at a singularity and take three (two) independent Jacobi fields 
along the geodesic. 
Roughly speaking, if the magnitude of a spacelike volume (area) element 
defined by the exterior product of the independent Jacobi fields becomes zero, 
then the singularity is called a strong curvature singularity.      

As a situation that the strong curvature singularity condition holds, 
we assume that the space-time is {\it maximal}~\cite{R}, 
whose definition we shall review below. 
Consider an affinely parameterized null geodesic $\lambda(v)$ 
whose tangent vector is $K^a$ and 
their two independent spacelike vorticity free Jacobi fields $Z_1$ and $Z_2$. 
Let $A$ be the area element defined by $Z_1 \wedge Z_2$ and 
introduce the function $z(v)$ defined by the relation $z^2 = A$. 
Then $z(v)$ satisfies the equation~\cite{T1} 
\begin{equation}
\frac{d^2z}{dv^2} 
  + \left(
          \frac{1}{2}R_{ab}K^{a}K^{b}+\sigma_{ab}\sigma^{ab} 
    \right) z =0,
\label{ray}
\end{equation}
where $R_{ab}$ is the Ricci tensor and $\sigma_{ab}$
is the shear of congruence of Jacobi fields along $\lambda$. 
Consider a null geodesic $\lambda:[0,a) \rightarrow M$ 
which is incomplete at the value $a$ of its affine parameter $v$ 
and generates an achronal set.  
It is said that $\lambda(v)$ satisfies the {\it inextendibility 
condition} if for some $v_1 \in (0,a)$ there exists a solution 
$z(v)$ of Eq.~(\ref{ray}) along $\lambda(v)$, with initial conditions: 
$z(v_1)=0$ and $\dot{z}(v_1)=1$, such that 
$\lim_{v \rightarrow a} \inf z(v)=0$.
As shown in Ref.~\cite{R}, $\lambda(v)$ cannot be extended beyond
a point $\lambda(a)$. This condition can be regarded as a strong
curvature singularity condition. 
It is also said that a space-time $(M,g)$ is {\it maximal} 
if the above inextendibility condition is satisfied 
for any incomplete null geodesic generating an achronal set in $(M,g)$.  

In Ref.~\cite{R} it is shown that there is no incomplete null geodesic 
generating an achronal set in a maximal space-time under some physical 
conditions. From this result we immediately have the following proposition. 

\begin{proposition}
\label{pr-r} 
Let $(M,g)$ be a maximal space-time on which the weak energy condition holds 
and $N$ a three-dimensional compact acausal set in $(M,g)$. 
If there is a point $p$ of $H^{+}(N)-N$ such that 
$J^{-}(p)\cap J^{+}(N)$ contains a future-inextendible 
timelike curve $\lambda$, 
then the null geodesic generator of an achronal set 
$[{\dot{J}}^{-}(p)\cap{\dot{J}}^{-}(\lambda)]\cap J^{+}(N)$ 
is future complete.  
\end{proposition}

In Theorem~\ref{th}, we showed that $\overline{V}$ is causally 
geodesically incomplete. 
Then, it is a possible case that ${\rm int} (V)$ is causally 
geodesically complete but $\overline{V}$ is not. 
Indeed, there are three places allowing this case in the proofs of our lemmas. 
%
%
The first place is in Lemma~\ref{pr-scc}, where we assumed that 
$\overline{\hat V}$ was null geodesically complete and 
thus the limit curve $\hat{\lambda}$ was complete in $\dot{\hat{V}}$. 
Hereafter, for simplicity, we assume that 
the strong causality condition is satisfied in $\hat{V}$ (condition~$2'$) 
because this assumption is reasonable for our considering space-time 
satisfying condition~2. 
Then one can prove Theorem~\ref{th} without assuming the completeness 
of $\hat{\lambda}$. 
%
%
The second place is where we assumed the future completeness 
of all the null geodesic generators of $\dot{J}^{-}(\hat{\gamma})$
in Lemma~\ref{lem-gamma2}. 
%
%
The third place is where we assumed 
the past completeness of $\hat{\beta}$ in Lemma~\ref{lem-gamma2}.  

Now, we can show the following theorem which states that 
if the considering finitely vicious space-time is maximal, 
the above case is impossible 
by combining Corollary~\ref{co-boundary} and Lemma~\ref{pr-B} below 
with Theorem~\ref{th}.  

\begin{theorem}
\label{th1}
If a finitely vicious space-time $(M,g)$ with a chronology violating 
set $V$ is maximal and satisfies conditions~$1,2',3$, 
then $V$ is causally geodesically incomplete. 
\end{theorem}  
%
Proposition~\ref{pr-r} leads to the following Corollary.  
\begin{corollary}
\label{co-boundary}
There is no future incomplete null geodesic generator of  
$\dot{\hat{V}}\cap{\dot{J}}^{-}(\hat{\gamma}) 
\cap J^+(\hat{K})$ 
in a maximal space-time $(\hat{M},\hat{g})$, 
where $\hat{\gamma}$ is the future inextendible
timelike curve obtained in Lemma~\ref{lem-gamma1}. 
\end{corollary}
\begin{proof}
Suppose that there was a future incomplete null geodesic generator 
$\hat{\lambda}$ of $\dot{\hat{V}}\cap
{\dot{J}}^{-}(\hat{\gamma})$. 
$\hat{\lambda}$ would also be a null geodesic generator of 
${\dot{J}}^{-}(\hat{p})\cap J^{+}(\hat{K})$ 
because 
$J^{-}(\hat{\gamma})\cap J^{+}(\hat{K})
\subset J^{-}(\hat{p})\cap J^{+}(\hat{K}) \subset \hat{V}$ 
by Lemma~\ref{lem-gamma1}, where $\hat{p}$ is 
the point defined in the proof of Lemma~\ref{lem-gamma1}. 
Consider a future-directed timelike curve $\hat{\mu}$ in 
$D^{+}(\hat{K})$ terminating at a point $\hat{s}$ of 
$\hat{\lambda}$. Take an infinite sequence of points $\hat{p}_n$ 
in ${\rm int} (D^{+}(\hat{K}))$ which converge to $\hat{p}$.   
Then ${\dot{J}}^{-}(\hat{p}_n)$ intersects $\hat{\mu}$ 
at a point $\hat{s}_n$ because ${\rm int}(D^{+}(\hat{K}))$ 
is causally simple. 
Denote an achronal null geodesic generator of 
${\dot{J}}^{-}(\hat{p}_n)\cap {\rm int}(D^{+}(\hat{K}))$ which 
connects $\hat{p}_n$ with $\hat{s}_n$ by $\hat{\xi}_n$. 
Since $\hat{p}$ and $\hat{s}$ are the limit points of 
$\{\hat{p}_n\}$ and $\{\hat{s}_n\}$ respectively, 
by Lemma 6.2.1 of Ref.~\cite{HE}  
$\{\hat{\xi}_n\}$ has two limit curves $\hat{\lambda}$ and $\hat{\xi}_p$, 
where $\hat{\xi}_p$ is a past-directed null geodesic generator of 
$H^{+}(\hat{K})$ through $\hat{p}$.    
$\hat{\xi}_p$ is past-inextendible by Lemma~\ref{lem-ch}. 
Therefore one can take a point $\hat{q}$ of $\hat{\xi}_p$ in the past 
of $\hat{p}$ and a convex normal neighborhood $\hat{G}$ of $\hat{q}$ 
with compact closure $\overline{\hat{G}}$ such that 
$\hat{\lambda}\cap\overline{\hat{G}}=\emptyset$. 
Since the space-time $(\hat{M}, \hat{g})$ is maximal, 
the future incomplete null geodesic 
$\hat{\lambda}:[0,a) \rightarrow \hat{M}$ would satisfy 
the inextendibility condition. 
Then one can get a contradiction by considering the Jacobi fields 
along $\hat{\xi}_n$ following the proof of the theorem by Rudnicki 
(see p.~57 of Ref.~\cite{R}). 
\end{proof}
\begin{lemma}
\label{pr-B}
There is a past incomplete null geodesic generator of 
${\dot{J}}^{+}(\hat{\beta})\cap\hat{V}$ 
if $\hat{\beta}$ is past incomplete, 
where $\hat{\beta}$ is the null geodesic generator of
$E^{-}(\hat{T})$ in Lemma~\ref{lem-gamma2}.  
\end{lemma}
\begin{proof}
Take an arbitrary point $\hat{w}$ of $\hat{\beta}$ and consider an 
infinite sequence of points $\hat{z}_l(\in\hat{V})$ on 
$\hat{\beta}_l$ which converges to $\hat{w}$, where 
$\hat{\beta}_l$ was defined in Lemma~\ref{lem-gamma2}. 
If one take a future-directed timelike curve $\hat{\rho}_l$ 
from $\hat{z}_l$, $\hat{\rho}_l$ intersects a point $\hat{w}_l$
on ${\dot{J}}^{+}(\hat{\beta})$. 
It is obvious that a subsequence of $\{\hat{w}_l \}$ converges to $\hat{w}$. 
Denote a future-directed null geodesic generator of 
${\dot{J}}^{+}(\hat{\beta})\cap J^{-}(\hat{T})$  
through a point $\hat{w}_l$ by $\hat{\kappa}_l$. 
Then one can take a subsequence of $\{\hat{\kappa_l}\}$ 
that converges to $\hat{\beta}$.

Suppose that the future endpoint $\hat{t}(\in\hat{T})$ of 
$\hat{\beta}$ was in $\dot{\hat{V}}$. 
Then, $\hat{t}$ was in $\pi^{-1}(\dot{K})$ or 
$\pi^{-1}(\dot{I}^{+}(q,M)\cap\dot{I}^{-}(q,M))$ by condition~1.   
Since $\hat{\beta}$ is the null geodesic generator of
$E^{-}(\hat{T})$, $\hat{\beta}$ (precisely, its extension) 
is also a null geodesic generator of 
${\dot{I}}^{-}(\hat{\gamma}) \cap \overline{\hat{V}}$, 
where $\hat{\gamma}$ is a future-inextendible timelike curve obtained 
in Lemma~\ref{lem-gamma1}. 
Thus, $\hat{\beta}$ was also a null geodesic generator of 
$\pi^{-1}(\dot{I}^{-}(q,M))$. 
Then $\hat{\beta}$ cannot remain in $\dot{\hat{V}}$ because $\hat{t}$
was in $\pi^{-1}(\dot{I}^{+}(q,M)\cap\dot{I}^{-}(q,M))$.     
This means that there was a point $\hat{b} (< \hat{t})$ of $\hat{\beta}$ 
in an arbitrary small neighborhood of $\hat{t}$ 
such that $\hat{b}\in (\hat{M}-\overline{\hat{V}})$. 
This is impossible because each $\hat{\beta}_l$ is in $\hat{V}$. 
Therefore $\hat{t}\in\hat{V}$ and one can take a  
neighborhood $\hat{U}_t$ of $\hat{t}$ 
such that $\overline{\hat{U}_t}$ is in $\hat{V}$. 
There is a natural number $l_o$ such that a subsequence of points 
$\hat{c}_{l'}$ of $\hat{\kappa}_{l'} \cap\hat{T}$ 
was in $\hat{U}_t$ for $l'\ge l_o$. 
Consider a causal curve which connects $\hat{z}_{l'}$ with 
$\hat{c}_{l'}$ through $\hat{w}_{l'}$. 
This causal curve is in $\hat{V}$ by Prop.~\ref{pr-ret}. 
Then each null geodesic generator $\hat{\kappa}_{l'}(l'\ge l_o)$
from $\hat{c}_{l'}$ to $\hat{w}_{l'}$ also is in $\hat{V}$. 
Since the point $\hat{w}$ is an arbitrary point of $\hat{\beta}$, 
there exists an infinite sequence of past-inextendible null geodesic curves 
$\hat{\kappa}_n$ of ${\dot{J}}^{+}(\hat{\beta})\cap 
J^{-}(\hat{T})$
in $\hat{V}$ which converge to $\hat{\beta}$.  

Since $(M,g)$ is the maximal space-time and $\hat{\beta}$ is a past 
incomplete null geodesic generating the achronal surface 
$E^{-}(\hat{T})$, $\hat{\beta}$  
satisfies the inextendibility condition. By this condition, 
$\lim_{v \rightarrow a}z(v)=0$ is satisfied along $\hat{\beta}$ in the past 
direction, where $v$ is an affine parameter of $\hat{\beta}$.     
Then there is a value $a'(a>a'>v_1>0)$ on $\hat{\beta}$ such that 
$\dot{z}(a')<0$ in the past direction. Consider a sequence of 
the Jacobi fields and the function $z_{n}$ along $\hat{\kappa}_n$ 
which satisfies Eq.~(\ref{ray}). 
By the continuity there is a natural number $n_1$ such that 
$\dot{z}_{n}(n\ge n_1)<0$ in the past direction. 
Thus $\hat{\kappa}_n(n\ge n_1)$ is also past incomplete in $\hat{V}$ 
by the energy condition 3, 
otherwise $\hat{\kappa}_n$ had pair conjugate points along $\hat{\kappa}_n$ 
which was impossible by the achronality of 
${\dot{J}}^{+}(\hat{\beta})\cap J^{-}(\hat{T})$. 
\end{proof}
\\

\begin{proofof}{theorem~\ref{th1}} 
By Theorem~\ref{th}, the assumption that $\overline{V}$ was causally 
geodesically complete is incorrect. 
Suppose that $V$ was causally geodesically complete 
but $\dot{V}$ was not. Then, recalling the proof of Theorem~\ref{th}, 
one could see that the null geodesic generator of 
${\dot{J}}^{-}(\hat{\gamma})$ in Lemma~\ref{lem-gamma2} 
was future incomplete or $\hat{\beta}$ in Lemma~\ref{lem-gamma2}
was past incomplete in $\dot{\hat{V}}$. By Corollary~\ref{co-boundary}      
the former case is impossible. 
By Lemma~\ref{pr-B} the latter case contradicts the supposition 
that $V$ was causally geodesically complete. 

\end{proofof}

\section{Chronology protection conjecture 
and causal feature of singularity } 
\label{IV}

In the previous section we showed that there exists an incomplete 
causal geodesic in $V$ under the some suitable conditions 
and the assumption that a finitely vicious space-time is maximal. 

\begin{figure}[htbp]
 \centerline{\epsfxsize=6.0cm \epsfbox{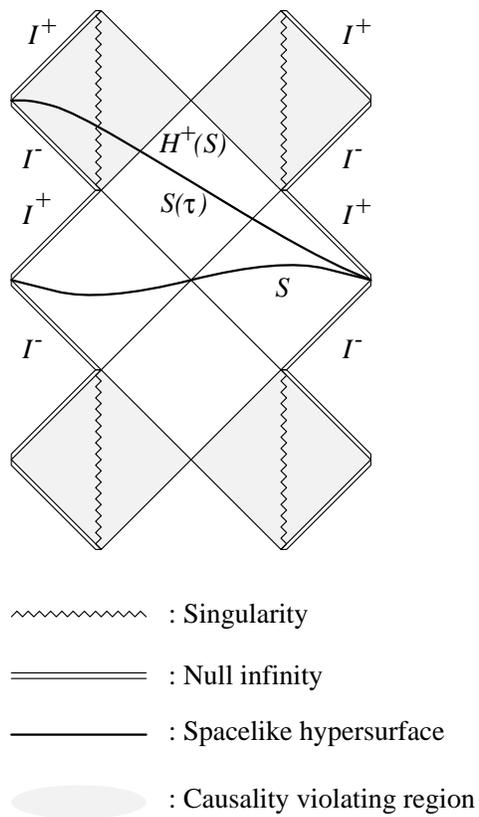}}
         \caption{
                  A conformal diagram of a maximally extended Kerr solution 
                  along the symmetry axis for non-extreme case. 
                  Singularities exist inside the causality 
                  violating-regions (the gray regions). 
                  Precisely, the zig-zag line in this figure describes 
                  the axis passing through, without intersecting, 
                  the ring of singularity 
                  rather than the singularity itself, 
                  since the only causal geodesics reaching 
                  the ring singularity are those in the equatorial plane. 
                  $S(\tau)$ is a locally spacelike hypersurface and 
                  $S(: = S(0))$ is a partial Cauchy surface. 
                  $H^+(S)$ is a future Cauchy horizon for $S$. 
                  }
               \protect
\label{Kerr}
\end{figure}

In this section we shall test the chronology protection conjecture 
by examining space-times considered in the previous sections, 
assuming further that 
the occurring singularities are physically realistic singularities. 
Here, we shall say that chronology is protected if $V=\emptyset$. 
We begin by inspecting exact solutions of the Einstein equation. 
One of the most well-known solutions containing both singularities 
and chronology violating sets is the maximally extended Kerr solution, 
which represents a rotating black hole. (See Ref.~\cite{HE}, for 
its metric and the maximal extension of the solution.) 
A conformal diagram of this solution is illustrated in Fig.~\ref{Kerr}. 
Chronology violating set is inside the inner horizons and 
the space-time is finitely vicious 
since a hypersurface slicing $\{S(\tau)\}$ can be taken so that the 
intersection of an $S(\tau)$ and the Cauchy horizon $H^+(S)$ 
for a partial Cauchy surface $S = S(0)$ is compact. 
As depicted in Fig.~\ref{Kerr}, this space-time has timelike singularities 
in the chronology violating set. It can be seen that 
the singularity is a real, scalar curvature singularity 
as the curvature scalar polynomial $R_{abcd}R^{abcd}$ diverges there. 

From the observation above, 
one may expect that even in more generic space-time 
an appearance of chronology violation of the finitely vicious type 
is accompanied by the formation of timelike (naked) singularities. 
In other words, from the standpoint of the chronology protection, 
one may conjecture that if all singularities are spacelike, 
chronology violation of the type cannot occur. 
To see that this conjecture is indeed true, 
we shall introduce the notion of a {\it non-naked singularity}, 
which is a generalization of a spacelike singularity observed, 
for example, in Schwarzschild space-time. 

\begin{definition}
\label{def1}
We shall say that a future (past) incomplete causal geodesic $\gamma$ 
terminates at a non-naked singularity if there is a point $p$ on $\gamma$ 
such that every future (past)-directed causal curve from $p$ terminates 
at scalar curvature singularities. 
\end{definition}  

A situation of occurrence of a non-naked singularity is illustrated 
in Fig.~\ref{NNS}. 
The non-naked singularity condition conforms to the belief that 
the formation of naked (timelike) singularities seems unlikely in a generic 
space-time in connection with the cosmic censorship conjecture~\cite{SCCC}. 
So this condition may be regarded as a seemingly reasonable condition 
for physically realistic singularities. 

It should be noted that this condition specifies the causal feature 
of a singularity in question. 
As depicted in Fig.~\ref{NNS}(a), non-naked singularities need not 
be ``everywhere spacelike'' and hence a space-time with this singularity 
needs not be globally hyperbolic even in the case of 
the absence of causality violation. 
Thus the condition is a quasi-global condition, so to speak.

Now, under this condition, we shall give a partial answer to the chronology 
protection conjecture by establishing the following theorem. 
\begin{theorem}
\label{pr-ch}
If every incomplete causal geodesic terminates at non-naked singularities, 
there is no maximal space-time $(M,g)$ which contains chronology violating 
set $V$ satisfying conditions~$ 1,2',3$. 
\end{theorem}   
\begin{proof}
Suppose that there were $V$ which satisfied conditions~$ 1,2',3$. 
By Theorem~\ref{th1} there is an incomplete causal geodesic $\gamma$ in $V$. 
Without loss of generality, one can consider $\gamma$ as a future-directed 
causal geodesic. 
Because all the occurring singularities are of the non-naked type, 
$\gamma$ has a point $p$ in the definition~\ref{def1}. 
On the other hand, there would be a closed timelike curve $\lambda$ 
through $p$ since $p$ was in the chronology violating set $V$. 
This contradicts the fact that all the future-directed causal curves 
from $p$ terminate at scalar curvature singularities. 
\end{proof} 

\begin{figure}[htbp]
 \centerline{\epsfxsize=9.5cm \epsfbox{FigNonNS.eps}}
    \caption{
             (a): A non-naked singularity. 
              $ \gamma$ is a future incomplete causal curve 
              which terminates at the non-naked singularity. 
             (b): A naked singularity, which is visible from the future null 
              infinity ${\cal I}^+$. 
             }
\protect
\label{NNS}
\end{figure}

\section{Concluding remarks}  
\label{V}  

As our main result we presented Theorem~\ref{pr-ch} which states     
that a physically reasonable class of chronology violations cannot 
arise if all occurring singularities are of the non-naked type. 
This supports the validity of the chronology protection conjecture. 

To prove this theorem, we first showed that there exists an incomplete 
causal geodesic in the chronology violating set $V$, if $V$ is non-empty.  
Although a number of studies have been made on the investigation of 
the relation between causality violation and singularities~\cite{T1,Kri,KA}, 
little has been known about the locations of the singularities. 
In Theorems~\ref{th} and~\ref{th1}, we succeeded in determining the locations 
by imposing some restrictions on the chronology violating set.    

The non-naked singularity condition in Theorem~\ref{pr-ch} implies 
that a version of the cosmic censorship conjecture~\cite{SCCC} is correct. 
Though this conjecture was proved in some seemingly physical cases~\cite{K}, 
whether or not this conjecture is correct in a more general context 
is still an open question.    
One may say that at least our result suggests that 
{\it the basic problems behind chronology protection strongly 
depend on the causal feature of occurring singularities.} 

As discussed in Ref.~\cite{THM}, in the case of 
spherically symmetric space-times one can discuss the causal feature 
of central singularity by using a quasi-local mass~\cite{QM}. 
We hope that this approach may provide a criterion for determining 
the causal feature of singularity in more generic cases.

\section*{Acknowledgments} 
We are grateful to I. Racz for useful discussion. 
We would also like to thank A. Hosoya and H. Ishihara for their continuous 
encouragement. The work is supported in part by the Japan Society 
for Promotion of Science (K.M.). 

\end{document}